\documentclass[fleqn,usenatbib]{mnras}
\usepackage[T1]{fontenc}
\usepackage{natbib}
\usepackage{psfig}
\usepackage{amsmath}
\usepackage{amssymb}
\usepackage{hyperref}
\def\be{\begin{eqnarray}}
\def\ee{\end{eqnarray}}
\usepackage{comment}
\usepackage{graphicx,subfigure}
\usepackage{appendix}
\usepackage{enumerate}
\usepackage{float}
\usepackage[flushleft]{threeparttable}

\title[A search for FRBs]{A targeted search for repeating fast radio bursts associated with gamma-ray bursts}

\author[N. T. Palliyaguru et al.]{
Nipuni~T.~Palliyaguru,$^{1,2}$,$^\dagger$\thanks{E-mail: nipuni.palliyaguru@ttu.edu}
Devansh Agarwal,$^{3,4}$
Golnoosh Golpayegani,$^{3,4,5}$ \newauthor
Ryan Lynch,$^{6,7}$ 
Duncan R.~Lorimer,$^{3,4}$
Benjamin Nguyen,$^{8}$
Alessandra Corsi$^1$ \newauthor
and Sarah Burke-Spolaor$^{3,4,9}$
\\
$^{1}$Department  of  Physics  and  Astronomy,  Texas  Tech  University,  Lubbock,  TX  79409-1051  (USA)\\
$^{2}$Arecibo Observatory, HC3 Box 53995, Arecibo, PR 00612\\
$^{3}$Department of Physics and Astronomy, West Virginia University, Morgantown, WV 26506-6315\\
$^{4}$Center for Gravitational Waves and Cosmology, West Virginia University, Chestnut Ridge Research Building, Morgantown, WV 26505\\
$^{5}$Department of Astronomy, University of California, Berkeley, 501 Campbell Hall \#3411, Berkeley, CA, 94720, USA\\
$^{6}$Green Bank Observatory, PO Box 2, Green Bank, WV 24944, USA\\
$^{7}$National Radio Astronomy Observatory, Charlottesville, VA 22903, USA\\
$^{8}$Department of Physics and Astronomy, Franklin and Marshall College, Lancaster, PA 17604, USA\\
$^{9}$CIFAR Azrieli Global Scholars program, CIFAR, Toronto, Canada
}
\date{Accepted XXX. Received YYY; in original form ZZZ}

\pubyear{2020}

\begin{document}
\label{firstpage}
\pagerange{\pageref{firstpage}--\pageref{lastpage}}
\maketitle

\begin{abstract}
The origin of fast radio bursts (FRBs) still remains a mystery, even with the increased number of discoveries in the last three years. Growing evidence suggests that some FRBs may originate from magnetars. Large, single-dish telescopes such as Arecibo Observatory (AO) and Green Bank Telescope (GBT) have the sensitivity to detect FRB~121102-like bursts at gigaparsec distances. Here we present searches using AO and GBT that aimed to find potential radio bursts at 11 sites of past $\gamma$--ray bursts that show evidence for the birth of a magnetar. We also performed a search towards GW170817, which has a merger remnant whose nature remains uncertain. We place $10\,\sigma$ fluence upper limits of $\approx 0.036$ Jy ms at 1.4 GHz and $\approx 0.063$ Jy ms at 4.5 GHz for AO data and fluence upper limits of $\approx 0.085$ Jy ms at 1.4 GHz and $\approx 0.098$ Jy ms at 1.9 GHz for GBT data, for a maximum pulse width of $\approx 42$ ms. The AO observations had sufficient sensitivity to detect any FRB of similar luminosity to the one recently detected from the Galactic magnetar SGR 1935+2154.  Assuming a Schechter function for the luminosity function of FRBs, we find that our non-detections favor a steep power--law index ($\alpha\lesssim-1.1$) and a large cut--off luminosity ($L_0 \gtrsim 10^{41}$ erg/s).
\end{abstract}


\section{Introduction}

Fast radio bursts \citep[FRBs;][]{lbm07,thornton13} are millisecond-duration radio pulses with large dispersion
measures, generally known to be of cosmological origin \citep[for recent reviews, see][]{phl19,cc19}, with the exception of FRBs from the Galactic magnetar SGR1935+2154 \citep{anderson2020,bmb20}.
Measurements to date imply total isotropic energies of the order of $\approx 10^{38}-10^{40}$~erg \citep{Law2017,z18,dgb15}, and high brightness temperatures that point to coherent emission processes. So far, there are over 100 known FRBs \footnote{\url{http://frbcat.org}} \citep{pbj16} with $\approx 20$ repeating \citep{ssh16,amiri2019,andersen19,fonseca20}.
Over 50 were found within the last three years by The Australian Square Kilometre Array Pathfinder \citep[ASKAP;][]{smb18}
 and The Canadian Hydrogen Intensity Mapping Experiment \citep[CHIME;][]{amiri18}.
 
Despite this rapid growth in observational results, the origin of FRBs is still uncertain. 
Theories proposed to explain the origin of FRBs include giant flares from magnetars \citep{pp10,kon14,mbm17}, giant pulses powered by spin-down from extragalactic neutron stars \citep[NS;][]{cw16}, produced from infalling asteroids to the pulsar's magnetosphere \citep{dww16}, collapse of massive NSs \citep{fr14} or compact binary mergers.
While there may be multiple classes of FRBs \citep{pz18}, the repeaters rule out catastrophic progenitors, at least for those particular objects.
The localization of the repeating FRB 121102 \citep{ssh16} to a low--metallicity dwarf galaxy \citep{tbc17,clw17}, started pointing to a NS origin for FRBs.
The more recent discovery of FRBs from the Galactic source SGR 1935+2154 \citep{anderson2020,bmb20} further confirms this.
The idea of a millisecond magnetar, a NS with millisecond birth period and a large magnetic field ($>10^{15}$ G), being responsible for the FRBs has been put forward since the discovery of FRBs \citep{pp10} and recently invoked to explain FRB121102 \citep{mbm17}.
Connections have also been made previously between FRBs and Soft Gamma Repeaters (SGRs), where the sudden release in magnetic energy that powers the SGR may also produce an FRB \citep{k16,anderson2020,bmb20}. 
In SGRs, magnetic reconnection produces a strong magnetic pulse that
propagates outwards and interacts with the surrounding gas, which could power a millisecond--duration burst in the radio band \citep{l14}.

Millisecond magnetars may be born in the core collapse of massive stars and/or in the merger of binary NSs \citep[e.g.,][]{usov92,dl98}.
The rotational energy of a NS is given by,
\begin{equation}
E_{\rm rot}\approx2\times10^{52} \rm erg \left(\frac{M}{1.4 \,\rm M_{\odot}}\right)\left(\frac{R}{10\,\rm km}\right)^2\left(\frac{P}{1\,\rm ms}\right)^{-1}.
\end{equation}
Rapid rotation (periods of milliseconds) gives sufficient rotational energy to power a $\gamma$--ray burst \citep{mgt11}.
Therefore, GRBs with intrinsic energy $<10^{52} \rm erg$ (which is the maximum rotational energy) may be powered by magnetars.
Energy injection from a rapidly rotating magnetar is often invoked to explain the energy in GRBs with a supernova (SN) connection \citep[hereafter called GRB-SNe,][]{mmw14}.
By comparing the kinetic energy of the associated SN and the $\gamma$--ray energy, \citet{mmw14} shows that a magnetar central engine likely powers all GRB--SNe.
In this scenario, the spin--down energy of a highly magnetized NS is deposited in the ejecta.

Evidence for the formation of NSs is also seen in some GRBs as a shallower-than-normal decay in the X-ray light curves \citep{zm01,cm09}.
For some of these GRBs, the central engine could be a magnetar, that did not immediately collapse to a BH, where the spin-down energy is deposited into the ejecta, giving rise to X-ray plateaus \citep{rom13}.

A merger that produces a stable NS that is rotationally supported may also quite possibly be a source of repetitive FRBs \citep{ytk17}.
The general picture concerning binary neutron star (BNS) mergers is that the merger produces either a BH or a long-lived NS depending on the
EOS and the maximum allowable mass for a NS, $M_{max}$, which could be more than 2.1 $M_{\odot}$ \citep{cromartie19}.
For BNS total mass of $\approx 1.3-1.6 M_{max}$, prompt collapse to a BH occurs, for masses $\lesssim 1.2 M_{max}$, a hypermassive NS is formed as an intermediate product, which then loses angular momentum and collapses to a BH \citep{mm17}, and for low-mass BNS, an indefinitely long-lived NS may be formed \citep{mm17}.

The FRB--GRB connection has also been discussed previously, with FRBs resulting from the collapse of a supra--massive NS to a BH \citep{z14}.
\citet{pwt14} searched for prompt radio emission coincident with GRBs, but these searches could have suffered from low sensitivity (SEFD~$>800$~Jy), whereas a typical FRB would have a flux density $\approx 0.5$ Jy.
Several other campaigns that followed up GRBs in search of FRBs have yielded non--detections \citep{Madison2019,Men2019,Hilmarson2020}.
However, \citet{bmg12} detected candidate single pulses from GRBs, \citet{dfm16}
identified a $\gamma$--ray transient in connection with an FRB, whose high isotropic $\gamma$--ray energy may be attributed to a magnetar origin, and \citet{Wang2020} identified a marginal association between FRB 171209 and GRB 110715A.
Adding to these, the presence of an SN remnant is one explanation for the large RM and the highly magnetized environment of the FRB121102 \citep{msh18}.
Intriguingly, \citet{nwb17} found that FRBs arising predominantly from repetitive sources, i.e. originating from magnetars,
occur at a rate of $10^4 \rm \, Gpc^{-3}$, consistent with the rates derived from observations.
It is also worthwhile pointing out that the birth rate of non--repeating FRBs ($\sim 2700\,\rm yr^{-1}\,Gpc^{-3}$) is consistent with the high end of the BNS merger rate \citep{nwb17,ytk17}.

Motivated by these ideas, we targeted several relatively nearby GRB sites and the BNS merger GW170817
in search of possible FRB signals.
A sensitive telescope like Arecibo should be able to detect FRB121102--like bursts of flux density $1.8$ mJy \citep{pwt14}
to a distance of up to $\approx4.8$ Gpc.
The details of our FRB search observations are presented in Section~\ref{obs}, data analysis in Section~\ref{da}, present results and place constraints on FRB progenitors in Section~\ref{discussion}, and we offer our conclusions in Section~\ref{conc}.

\section{Observations}
\label{obs}

Our target list contains 11 GRBs visible in the Arecibo Observatory (AO) declination range and the BNS merger GW~170817, which was observed with the Green Bank Telescope (GBT). Out of the 11 GRBs, there are six long GRB (lGRB) with SN associations, and five GRBs (one lGRB and four short GRBs; sGRBs) that exhibit plateaus in the X--ray light curve.
The target list and their properties
are given in Table~\ref{tb:GRBlist1}. 
All GRBs have been localized to a region $<2.5$ arcsec by high energy observations, which allows single-dish telescopes with a field-of-view (FoV) of a few arcminutes to cover the entire localization error region at full sensitivity.

\begin{figure}
    \begin{center}
\includegraphics[width=\columnwidth]{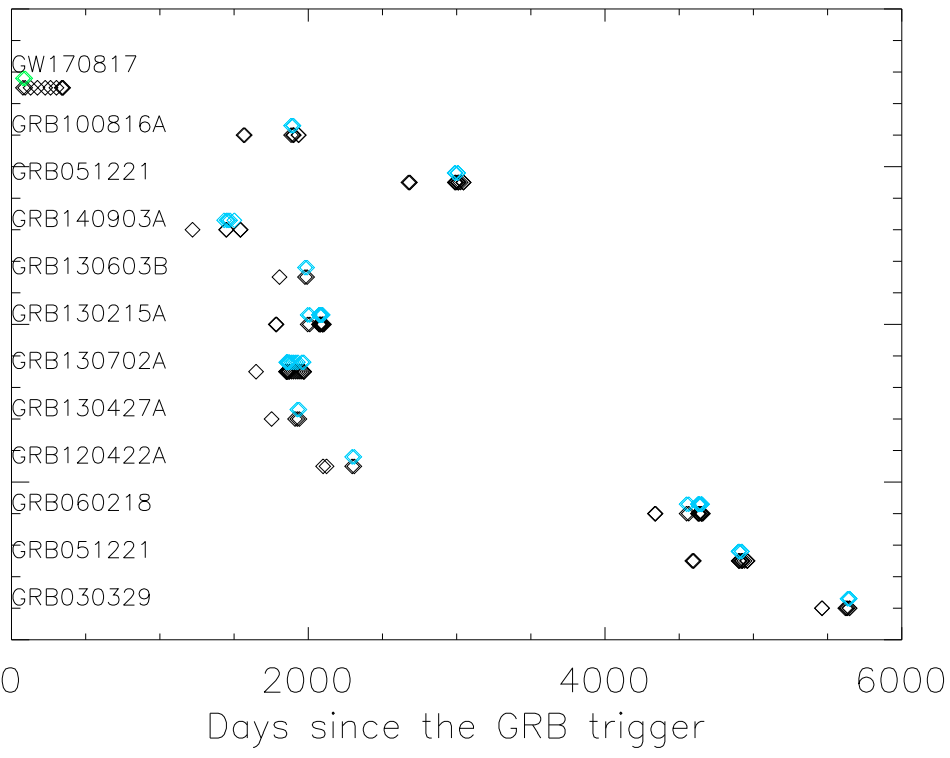}
\caption{
The timescale of our radio observations for the sources that are listed in Table~\ref{tb:GRBlist1} since the GRB trigger. All observations were made within a time span of 412 days, between the dates of November 2, 2017 and December 19, 2018. Our targets are arbitrarily offset along the y-axis for clarity. Black, blue and green diamonds represent 1.4~GHz, 4.5~GHz and 2~GHz observations, respectively. GRB~130702A was monitored $\sim$ once a month in order to verify the origin of an excess dispersion measure seen on the first observation.
\label{fig:obs}
}
\end{center}
\end{figure}

Full stokes (or polarization self-and cross-products), 8--bit, high time resolution spectra were recorded at both telescopes.
A known pulsar was observed at the start of each observation session to verify the status of instruments.
Table~\ref{tb:obsinfo} lists information about the observation setup for AO and GBT.
The dates since the GRB trigger, on which the observations were conducted, are shown in Figure~\ref{fig:obs}.

\begin{table*}
\begin{center}\begin{footnotesize}
\caption{The list of targets.
Name, redshift, distance, RA, DEC, galactic coordinates ($l$ and $b$), expected DM, GRB duration ($\rm T_{90}$), and the minimum detectable luminosity ($L_{\rm min}$) corresponding to S/N$=10$ at 1.4~GHz.
}
\label{tb:GRBlist1}
\resizebox{18cm}{!}{
\begin{tabular}{lllllllllllllll}
\hline
Name&Redshift&Distance&RA&DEC&$l$&$b$&Expected DM&Duration& $ L_{\rm min}$\\
&&Mpc&(deg)&(deg)&(deg)&(deg)&($\rm cm^{-3}$ pc)&(sec)& erg/s\\
\hline
GRB\,030329/SN2003dh&  0.169&812& 10:44:50.03& 21:31:18.15& 216.98& 60.69&  264.32 & 25 & $1.72\times10^{40}$     \\
GRB\,060218/SN2006aj&  0.033&145&  03:21:39.69& 16:52:01.6& 166.92&  -32.88& 173.08 & 2100 &$5.5\times10^{38}$   \\
GRB\,120422A/SN2012bz& 0.280&1434& 09:07:38.42& 14:01:06.0& 215.22&  36.437& 413.64 & 5.35       &$5.38\times10^{40}$ \\
GRB\,130427A/SN2013cq& 0.340&1795& 11:32:32.63& 27:41:51.7& 206.51&   72.50& 472.37 & 162.83      &$8.44\times10^{40}$  \\
GRB\,130702A/SN2013dx& 0.145&687&  14:29:14.78& 15:46:26.1&  11.35&  64.63&  241.17 & 59 &$1.23\times10^{40}$  \\
GRB\,130215A/SN2013ez& 0.597&3508& 02:53:56.6 & 13:23:13.2& 163.07&  -39.76& 991.80 & 65.7  & $3.22\times10^{41}$   \\
GRB\,130603B&          0.356&1894& 11:28:48.15& 17:04:16.9& 236.42&  68.42 & 498.47 & 0.18       &$9.39\times10^{40}$ \\
GRB\,140903A&          0.351&1863& 15:52:03.27& 27:36:09.4& 44.39&   50.12&  496.38 & 0.30      &$9.09\times10^{40}$ \\
GRB\,051221&   0.547&2868 &  21:54:48.71& 16:53:28.2& 73.54&   -28.58&786.02  & 1.4  &$2.15\times10^{41}$ \\
GRB\,100816A&  0.803&4529 &  23:26:57.62& 26:34:43.9 & 100.44& 32.57&1396.40 & 2.9 &$5.37\times10^{41}$                        \\
GRB\,130831A&  0.479&2459 &  23:54:29.91& 29:25:47.6 &  08.33& -31.82&663.84 &  32.5      &$1.58\times10^{41}$            \\
GW170817 & 0.0098 & 42 &13:09:48.089& -23:22:53.35 & 308.37&        39.29 7 & 77.27 & 2.0& $1.06\times10^{38}$  \\
\hline
\end{tabular}
}
\begin{tablenotes}
      \small
      \item References: \citet{smg03}, \citet{cpp14},  \citet{mmw14},  \citet{dpm15}, \citet{abbot1} and \url{https://swift.gsfc.nasa.gov/archive/}.
\url{https://swift.gsfc.nasa.gov/archive/grb_table/}
    \end{tablenotes}
\end{footnotesize}\end{center}\end{table*}

Arecibo observations were carried out between 2017 December 12:50~UTC and 2018 December 19:55~UTC. 
Arecibo targets were observed for $\sim 0.6$ hours on each epoch.
A total of 114 hours of observations were obtained on all GRBs, with 1--21 hours on each target at each frequency depending on the LST availability.
We note that all of the sources are away from the Galactic plane.
Data were recorded using the Puerto Rico Ultimate Pulsar Processing Instrument (PUPPI) at center frequencies of 1380~MHz and 4.5 GHz with a bandwidth of $\approx600$ MHz with 512 frequency channels
and sampled at 
$40.96\,\mu\rm s$.

\begin{table}
\begin{center}\begin{footnotesize}
\caption{Summary of observational setup. Telescope name, center frequency ($f_c$), System temperature ($T_{\rm sys}$), Telescope gain, Bandwidth, number of channels ($N_{\rm chan}$), sampling time ($t_{\rm samp}$) and the FoV.
}
\label{tb:obsinfo}
\resizebox{\columnwidth}{!}{
\begin{tabular}{lllllllllllllll}
\hline
Name &$f_c$ &$T_{\rm sys}$ & Gain & BW &$N_{\rm chan}$&$t_{\rm samp}$ & FoV\\
Telescope &(MHz)&K&(K/Jy)& (MHz) && $\mu$s & $\rm deg^2$\\
\hline
AO & 1.4 & $\approx$30 & 8 & 600 & 512 & 40.96 & $9.5\times10^{-3}$\\
AO & 4.5 &$\approx$30 & 4 & 800 & 512 & 40.96 & $7.8\times10^{-4}$ \\
GBT & 1.4 & $\approx$20 & 2.0 & 800 & 512 & 10.24 & $1.8\times10^{-2}$  \\
GBT & 1.9 & $\approx$22 & 1.9 & 800 & 512 & 10.24 &$8.7\times10^{-3}$\\
\hline
\label{tb:GRBlist2}
\end{tabular}  
}
\begin{tablenotes}
      \small
      \item $ T_{\rm sys}$, Gain and FoV values for AO and GBT are obtained from  \url{http://www.naic.edu/~astro/RXstatus/rcvrtabz.shtml} and 
\url{https://science.nrao.edu/facilities/gbt/proposing/GBTpg.pdf} respectively.
    \end{tablenotes}
\end{footnotesize}\end{center}
\end{table}



GBT observations were carried out between 2017 November 02:41~UTC and 2018, July 31:02~UTC on 10 epochs.
GW~170817 was observed for $\sim 1$ hour at each frequency during each observation.
Since low-frequency emission may be self-absorbed by the post-GRB ejecta at early epochs, observations were conducted at 1.4 and 1.9 GHz on the first two epochs and afterwards only at 1.4 GHz on the following epochs.
GBT data were recorded on the GUPPI spectrometer with a sampling time of $10.24\,\mu\rm s$.
To minimize any intrachannel dispersive smearing to that caused by the difference between the true and estimated DM, data were semi-coherently dedispersed at a DM of 80 cm$^{-3}$~pc.
This DM is an estimate corresponding to DM = DM$_{\rm MW}$ + DM$_{\rm IGM}$ + DM$_{\rm H}$, 
where $\rm DM_{MW}=35 \, \rm cm^{-3}$~pc is the contribution from the Milky Way \citep{ymw17}, 
$\rm DM_{\rm IGM}=7 \, cm^{-3}$~pc, is the intergalactic medium \citep{iok03}, and DM$_{\rm H} =35\, \rm cm^{-3}$~pc is the host galaxy (assuming a host like the MW).
We use DM$_{\rm H} = 35$ cm$^{-3}$~pc, which is the approximate DM contribution from the Milky Way for a line of sight out of the plane.  


\section{Data Analysis}
\label{da}

Data from both telescopes were processed with the  GPU accelerated pipeline \textsc{Heimdall}\footnote{\url{https://sourceforge.net/projects/heimdall-astro/}} and the PRESTO analysis package \citep{r01}.

\begin{figure}
    \begin{center}
    \hspace{-1.cm}
\includegraphics[width=\columnwidth,trim={ 4.5cm 5.0cm 4.5cm 4.4cm},clip]{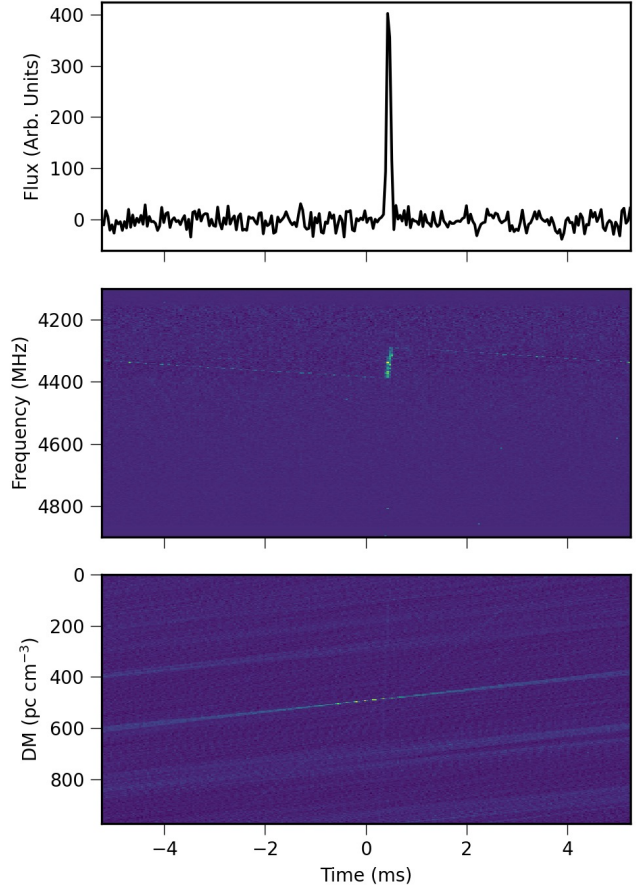} 
\caption{
Candidate burst (chirped RFI) from the single--pulse search pipeline in FETCH from GRB\,051221 with the brightest signal at $\rm DM=486.4\,\rm pc\,cm^{-3}$. The signal is shown as flux density vs time (top), frequency vs time (middle) and DM vs time (bottom).
\label{fig:rfi}
}
\end{center}
\end{figure}

We search the data at the recorded frequency and time resolution using a GPU accelerated pipeline \textsc{Heimdall}. The data were dedispersed at 1--7000 $\rm pc\,cm^{-3}$ and were searched for pulse widths of 40.96~$\mu$s--41.93~ms with the increments in the power of two. This generated 17,672 candidates above S/N of 6. These candidates were then fed to convolutional neural network \textsc{FETCH}\footnote{\url{https://github.com/devanshkv/fetch}} to classify candidates between radio frequency interference (RFI) and potential FRB candidates \citep{agarwal20}.
We use model \textsc{a} with a probability threshold of 0.5 for the candidate classification. \textsc{FETCH} labelled 425 candidates as positives. These were inspected manually, 68 candidates were single pulses from the above mentioned test pulsars. The rest of the candidates were false positives due to a nearby airport radar.
Figure~\ref{fig:rfi} shows an example candidate appearing at a non--zero DM, due to chirped RFI from radar at 4.5~GHz. The appearance of the signal within a narrow band of  4250-4350~MHz confirms the non--astrophysical nature of the signal.

Within PRESTO, RFI excision was done using the tool \texttt{rfifind}, which creates a mask in which the affected frequency channels and time chunks are replaced by median values.  
The data were 
referenced to the Solar System barycentre
and de-dispersed at 1000 trial DMs ranging from 0 to 1000~cm$^{-3}$~pc. 
For GRB130215A and GRB100816A, since the expected DM is $>1000\, \rm cm^{-3} pc$ (as listed in Table~\ref{tb:GRBlist1}), we search 333 trial DMs ranging from 1000 to 2000 $\rm cm^{-3} pc$.
The dispersion smearing  within a channel of $\Delta\nu=800/512$ MHz at the lowest frequency of 0.98~GHz from this step size is 8.3$\mu$s~DM~$\Delta\nu\nu^{-3}\approx 13.7 \mu$s, which is negligible.

Significant peaks ($>10~\sigma$) were searched for, in the de-dispersed time series using the single--pulse pipeline in PRESTO.
Each DM vs time plot was visually inspected for real bursts, 
and those that peaked at a DM of zero were considered as RFI.
Candidates that appeared at non-zero DM were reprocessed with \texttt{dspsr} \citep{vb11} using the DM and time output by the single--pulse search pipeline in PRESTO, and plotted using PSRCHIVE plotting routines \citep{hsm04}.
Candidate bursts detected in the single--pulse search were classified as RFI upon further examination.

\begin{table}
\begin{center}\begin{footnotesize}
\caption{GRB name, frequency, total time on the source, the number of epochs observed and the average time duration of each observation.
}
\begin{tabular}{lllllllllllllll}
\hline
Name &Frequency & Total & Number & Average \\
& & Time & & Time\\
&(MHz)&(hours)&&(hours)\\
\hline
GRB\,030329& 1380 & 4.5& 7 &0.65\\
& 4500 & 1.7& 4& 0.42\\
GRB\,051221& 1380 &14.7&15 &0.98\\
& 4500 & 2.5& 5& 0.51\\
GRB\,060218& 1380 & 8.7&16 &0.55\\
& 4500 & 6.9&13& 0.53\\
GRB\,120422A& 1380 & 3.5& 4 &0.88\\
& 4500 & 1.2& 2& 0.60\\
GRB\,130427A& 1380 & 2.7& 4 &0.67\\
& 4500 & 1.0& 2& 0.51\\
GRB\,130702A& 1380 &13.7&21 &0.65\\
& 4500 & 6.9&15& 0.46\\
GRB\,130215A& 1380 &12.1&19 &0.64\\
& 4500 & 7.5&12& 0.63\\
GRB\,130603B& 1380 & 2.3& 3 &0.76\\
& 4500 & 1.0& 2& 0.50\\
GRB\,140903A& 1380 & 4.1& 6 &0.68\\
& 4500 & 3.1& 6& 0.51\\
GRB\,051221& 1380 &14.7&15 &0.98\\
& 4500 & 2.5& 5& 0.51\\
GRB\,100816A& 1380 & 5.4& 8 &0.67\\
& 4500 & 1.3& 3& 0.44\\
GW\,170817& 1399 &8.7&10 &0.86\\
& 1999 & 1.8& 2& 0.91\\

\hline
\label{tb:GRBlist2}
\end{tabular}
\end{footnotesize}\end{center}
\end{table}

\section{Discussion}
\label{discussion}

In this section, we discuss the timescales for radiation to escape the GRB ejecta, the possibility of detecting FRBs from magnetars based on the expected flux density and the constraints placed on the luminosity function based on non--detections.

\subsection{Detectability of a repeating FRB}
\label{det}

The environment of the burst/merger site is an important consideration when determining the timescales for radio emission to escape. The free--free optical depth for the (Oxygen dominated) ejecta would reach $\tau_{\rm eff}=1$, on a timescale
\begin{equation}
    t=10\times\left(93\,f_{\rm ion}^2\nu_{\rm GHz}^{-2} T_4^{-3/2} M_{10}^2 t_{10} v_9^{-5}\right)^{5}~{\rm yr},
\end{equation}
after the explosion \citep{mbm17},
where $f_{\rm ion}$ is the ionized fraction of the ejecta, $\nu_{\rm GHz}$ is the observing frequency, $M_{10}$ is the ejecta mass in units of $10M_{\odot}$ and $T_4$ is the ejecta temperature in units of $10^4$ K.
Assuming ejecta masses of $\sim 10 M_{\odot}$, $f_{\rm ion}=0.4$, ejecta velocities $\sim 10^4$km~s$^{-1}$, $T_4=1$ radio emission may escape after 6 and 3 yr after the explosion at frequencies 1~GHz and 4.5~GHz respectively.
With relatively smaller ejecta masses in mergers (than in SNe), ejecta will be transparent to $\sim 1$~GHz emission on a timescale of several months \citep{mbm17}.
This could be as soon as $\sim$three months for ejecta mass of $\sim 0.001\,M_{\odot}$.
Our search observations were carried out $\approx$~3 months--15~yr after the explosion.

\begin{figure}
    \begin{center}
\includegraphics[width=\columnwidth,trim={ 1.0cm 12.0cm 2.0cm 2.8cm},clip]{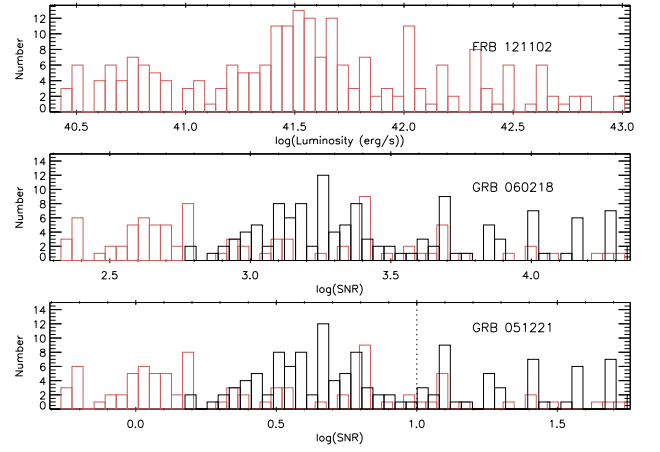} 
\caption{
Luminosity distribution of FRB\,121102 bursts (top), S/N histograms of GRB\,060218 (middle) and GRB\,051221 (bottom). The S/N histograms show the S/N at which FRB~121102--like bursts would be detected at given GRB distances. The black dotted vertical line corresponds to $S/N=10$. The red and black S/N histograms represent bursts at 1.4~GHz and 4.5~GHz respectively.
}
\label{fig:hist}
\end{center}
\end{figure}

Measured flux densities of bursts from the repeating FRB~121102 were scaled to the distance of each GRB to estimate the expected flux density if a repeater--like source resided in the GRB site.
Assuming radiometer noise limitations for each burst, the signal-to-noise ratio
\begin{equation}
    {\rm S/N} = \frac{F\,G\sqrt{N_p\Delta \nu}}{\beta T_{\rm sys} \sqrt{w}},
\end{equation}
where $F$ is the expected fluence given by $F=S\, w$ where $S$ is the flux density,  $w=1.0$~ms is the pulse width, $T_{\rm sys}$ is the system temperature, $G$ is the telescope gain (the numbers given in Section~\ref{obs}),  $\Delta \nu$ is the bandwidth, $\beta=1.07$ accounts for digitization loss factors, and $N_p=2$ is the number of polarizations \citep{rlb16}.
From the radiometer equation, the minimum flux density corresponding to S/N~$=10$ for the AO setup is $S_{\rm min}\approx36 \, \rm mJy$ at 1.4~GHz and $S_{\rm min}\approx63 \, \rm mJy$ at 4.5~GHz.
The minimum flux density for GBT is $S_{\rm min}\approx 84 \, \rm mJy$ at 1.4~GHz and $S_{\rm min}\approx98 \, \rm mJy$ at 1.9~GHz.
Figure~\ref{fig:hist} shows the luminosity distribution and the S/N of 224 bursts at $1.4$ GHz and $4-6$ GHz from FRB~121102 \citep{pwt14,Hardy2017,Spitler2018,Zhang2018,msh18,Magic2018,Gourdji2019,Hessels2019}. The S/N histograms for bursts when scaled to the distances of two example GRBs from our sample are also shown at 1.4~GHz and 4.5~GHz.
The S/N histograms are created by scaling flux density of FRB~121102 bursts to the distances of the GRBs and calculating the ratio between the expected flux density and minimum flux density $S_{\rm min}$.
If magnetars emit FRB121102--like bursts, Arecibo should be able to detect the brightest bursts of luminosity $\approx9\times10^{42} \, \rm erg/s$ (flux density of 1.8~Jy at 4.5 GHz) at GRB distances up to 4.8~Gpc.

\subsection{Luminosity function}

In this section, we attempt to place an upper limit on the FRB rate and constrain the FRB luminosity function parameters based on the non--detection of FRBs in our data. 
The FRB luminosity function may be expressed by the Schechter function (Schechter 1976) where the event rate density  per luminosity interval $d L$ is given by
\begin{equation}
\phi(L)\,dL=\phi^{*}\left(L/L_0\right)^{\alpha} e^{-L/L_0}d(L/L_0),
\end{equation}
where $\phi^*$ is a reference event rate density, $\alpha$ is the power law exponent, and $L_0$ is the cut--off luminosity. 

Following \citet{Luo20}, the event rate above a minimum luminosity $L_{min}$ is then given by,
\begin{eqnarray}
R(>L_{\rm min}) &=& \int_{L_{min}/L_0}^\infty \phi^{*}\left(L/L_0\right)^{\alpha} e^{\left(-L/L_0\right)} d\left(\frac{L}{L_0}\right) \nonumber \\
&=&R_0\,\Gamma_i\left(\alpha+1,\frac{L_{\rm min}}{L_0}\right),
\label{eq:rate}
\end{eqnarray}
 where 
 $R_0$ is the all--sky event rate in $\rm sky^{-1}\,day^{-1}$, 
 $\Gamma_i$ is the incomplete gamma function. 
Here we have replaced the volumetric rate by the all--sky rate. 
The minimum detectable luminosity at the $i^{\rm th}$ GRB, at a distance $D_i$, is calculated as
$L_{{\rm min},i}= 4\pi\,\Delta\nu\,S_{\rm min} D_i^2$.
Here, $\Delta\nu$ is the bandwidth and $S_{\rm min}$ is the minimum detectable flux density, calculated from the radiometer equation.
The minimum flux density corresponding to S/N~$=10$ for AO and GBT are given in Section~\ref{da}.
Table~\ref{tb:GRBlist1}
lists $L_{\rm min}$ values for each GRB at 1.4~GHz.
The minimum luminosity of the AO sources at 4.5 GHz is $\approx 2.3$ times larger than at 1.4 GHz. 
The minimum luminosity for GW 170817 at 1.9 GHz is $1.64\times10^{38} \, \rm erg\, s^{-1}$.

If the $i^{th}$ GRB site was searched for $T_i$ days, the expected number of pulses for at that GRB is, 
\begin{equation}
    n_i =  \left(\frac{R\,T_i\,\Omega}{41253\, \rm deg^2}\right).
    \label{eq:npulses}
\end{equation}

Here, $\Omega$ is the FoV of the telescope (listed in Table~\ref{tb:obsinfo}) and $T_i$ is the observation time on the $i^{th}$ source.
Therefore, assuming a Poisson distribution of pulses, the probability of observing zero pulses in the $i^{\rm th}$ GRB is, $p_i = e^{-n_i}$.
The likelihood of not detecting any GRBs in the entire sample, $\mathcal{L}$, is the product of all the probabilities, i.e.
\begin{equation}
\mathcal{L} = \displaystyle \prod_{i=1}^{N} p_i.
\end{equation}
By summing individual logarithms, we see that
\begin{equation}
    \log (\mathcal{L})=-R_0\sum_{i=1}^{N}\Gamma_i\left(\alpha+1,\frac{L_{{\rm min},i}}{L_0}\right)\, T_i \frac{\Omega}{41253 {\, \rm deg}^2}.
    \label{eq:likelihood}
\end{equation}

\begin{figure}
    \begin{center}
    \hspace{-1.cm}
\includegraphics[width=\columnwidth,trim={1.0cm 12.5cm 2.5cm 2.0cm},clip]{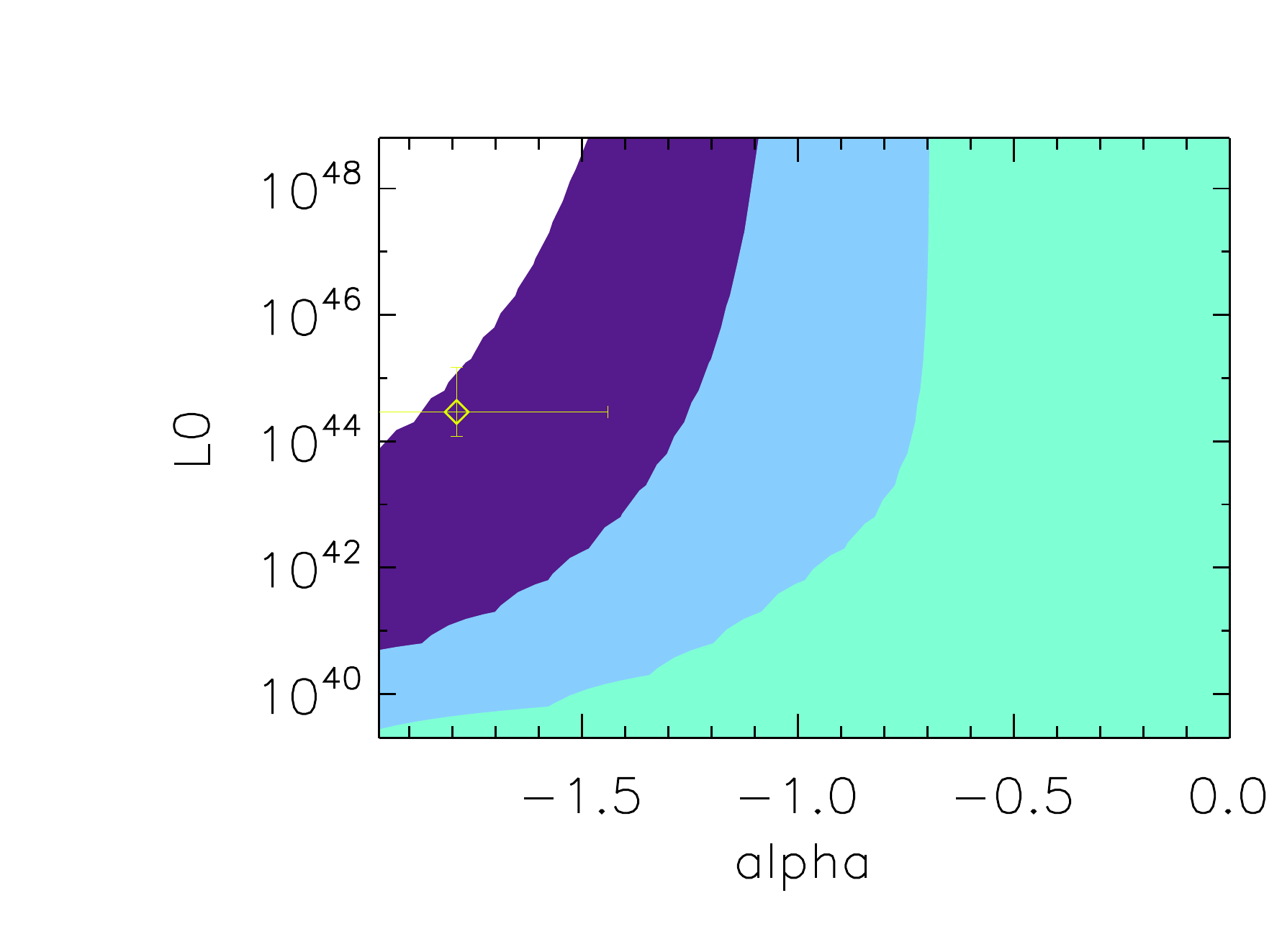}
\caption{
$R_0$ vs $L_0$ and $\alpha$
for Schechter function for 95\% nondetection probability.
The white, purple, light blue, and aqua green regions correspond to
 $R_0 \lesssim 0.64 \, \rm sky^{-1}\, day^{-1}$, between  $0.64$ and $937 \, \rm sky^{-1}\, day^{-1}$, between 937 and $1.8\times10^4 \, \rm sky^{-1}\, day^{-1}$, 
 and $\gtrsim 1.8\times10^4 \, \rm sky^{-1}\, day^{-1}$, respectively.
\citet{Luo20} LF parameters of $L_0=2.9^{+11.9}_{-1.7}\times10^{44} \rm \, erg/s$ and $\alpha=-1.79^{+0.35}_{-0.31}$ are marked by a yellow diamond.}
   \label{fig:ratevsalpha_l0}
\end{center}
\end{figure}

For a given non--detection probability, we can place an upper limit on the all--sky rate and place constraints on the parameters of the luminosity function using Equation~\ref{eq:likelihood}.
Figure~\ref{fig:ratevsalpha_l0} shows $R_0$ vs $\alpha$ and $L_0$ values for $95\%$ probability.
We have combined all data from both AO and GBT.

A millisecond magnetar  birth rate of $\approx 170 ~\rm Gpc^{-3}\,yr^{-1}$ 
and a core--collapse supernova rate of $\approx 2.5\times 10 ^5 ~\rm Gpc^{-3}\,yr^{-1}$ \citep{nwb17}, converted to an all--sky rate 0.64 and 937 $\rm sky^{-1}\,day^{-1}$ respectively, are marked by the purple region.
To convert the volumetric rate to an all--sky rate we assume that the current AO setup can detect an FRB of 0.19 mJy at 1.4 GHz \citep[mean flux density of FRB 121102;][]{pwt14} up to a distance of $\approx 2.2$ Gpc 
with $\rm S/N=10$). 
We also assume a duty cycle of $\eta\approx0.1$ and a beaming factor of $\xi\approx0.3$.
The sky rate for non--repeating FRBs, scaled to our observations such that $R_0\approx1.8\times10^{4}\rm~sky^{-1}$~day$^{-1}$ above 36 mJy, is marked by the blue region. 
Here we use the expression for the all--sky rate 
\begin{equation}
    R(>S)=R_0\left(\frac{S}{\rm Jy}\right)^{\alpha^\prime},
\end{equation}
where $S$ is the minimum flux density, $R_0=1140\rm~sky^{-1}$~day$^{-1}$, is the reference rate at a flux density of 1 Jy  and $\alpha^\prime=-0.83$ is the source count index \citep{als20}
from the log N-log S relation \citep{lvl17}.
To be consistent with the expected rates of millisecond magnetar formation, $L_0$ should be higher for smaller $\alpha$ (purple region). In general $\alpha\lesssim-1.1$ and $L_0 \gtrsim10^{41}$~erg~s$^{-1}$.

\citet{Luo20} finds from real FRBs luminosity function parameters $L_0=2.9^{+11.9}_{-1.7}\times10^{44}  \rm \, erg \, s^{-1}$, $\alpha=-1.79^{+0.31}_{-0.35}$ and a volumetric rate of $\phi^*=339 \,\rm Gpc^{-3}\,yr^{-1}$.
This corresponds to an all--sky rate of $R_0\approx 42.4 \rm \, sky^{-1}\,day^{-1}$ at 1.4 GHz. 
From Figure~\ref{fig:ratevsalpha_l0}, and Equation~\ref{eq:likelihood} we find that for $L_0=2.9\times10^{44} \rm \, erg s^{-1}$ and $\alpha=-1.79$ gives $R_0\approx 2.4 \,\rm sky^{-1}\,day^{-1} $, above a flux density of 36 mJy.
A flat spectrum for FRBs is assumed here.
We further note that the all--sky FRB rate for repeaters will be a fraction of this rate.

Recent studies have shown that FRB 121102 shows a periodicity and clustering and therefore the burst distribution may be better described by a Weibull distribution than a Poisson distribution \citep{Oppermann2018}.
However, this may be the effect of a few strongly clustered bursts and the burst distribution may still be Poissonian \citep{Cruces2020}.
The non--detection probability for a Weibull distribution is given by \citep{Oppermann2018}
\begin{equation}
    p=\frac{\Gamma(1/k)\,\Gamma_i(1/k,(T_{i}r\,\Gamma(1+1/k)^k))}{k\Gamma(1+1/k)}
\end{equation}
where $k$ is the shape parameter
which describes the degree of clustering,
$r$ is the burst rate, $T_i$ is the observation time, $\Gamma$ is the Gamma function, $\Gamma_i$ is the incomplete Gamma function.
For $k=1$ the Weibull distribution becomes a Poisson distribution and for k<1 clustering with small intervals is favoured \citep{Oppermann2018}.
Considering a burst rate of $r=5.7 \, \rm day^{-1}$ \citep{Oppermann2018} and scaling the rate to the distance of each GRB, and taking the product of probabilities for each GRB,  the total non--detection probability for AO observations is $3.9\times10^{-6}$ for a Poisson process.
For a Weibull distribution with $k=0.34$ \citep{Oppermann2018}, the non--detection probability is $1.8\times10^{-5}$.
Here we ignore GRB 060218, since it has a high rate and hence a very small non--detection probability.

\section{Conclusion}
\label{conc}
We conducted a single--pulse search for FRBs from 12 well--localized targets that show evidence for magnetar formation.
The target list includes six GRB-SNe, four sGRBs, one lGRB without a SN association, and GW170817, for which the merger remnant is undetermined.
These searches were conducted $\sim 3$ months--15 years after the explosion.
We show that large single--dish telescopes are well suited to detect FRBs from such extragalactic targets at Gpc distances. Our searches resulted in candidates that were confirmed to be either RFI or single pulses from the known test pulsars.
Our constraints on the FRB luminosity function parameters, based on non--detection, are consistent with published values. 

The detection of a late--time FRB signal from a GRB site would undoubtedly be the smoking gun signature of magnetar birth which would have a tremendous scientific impact with vast implications for fundamental physics and cosmology for this decade \citep{lmp19}.
New wide--field radio telescopes have more than doubled the number of FRBs over the last two years.
However, even with the increased number of bursts in the last two years, mechanisms that produce FRBs remain a mystery.
Determining the observing cadence remains one of the main challenges in targeted searches.
If FRBs are indeed related to explosive events, a better understanding of the emission process and the environment of the explosion will help determine factors such as time for radiation to escape and thereby an observing cadence for future targeted searches. 
Novel techniques to catch possible radio bursts from gravitational wave counterparts are emerging (Clancy et al. 2019).
Better algorithms that reduce the number of candidates and distinguish between RFI and real transients are also in place.
Furthermore, even though radio telescopes with large fields--of--view is dominating FRB searches, sensitive single--dish telescopes will continue to play a crucial role in follow--up searches at targeted locations.

\section*{Acknowledgements}
NTP thanks the Phill Perrillat, Hector Hernandez and other Arecibo Observatory staff for data quality checks, scheduling and support with observing.
We thank Akshaya Rane, Jayanth Chennamangalam, Andrew Seymour and Scott Ransom for help with observations and data processing issues.
The Arecibo Observatory is a facility of the National Science Foundation operated under cooperative agreement by the University of Central Florida in
alliance with Yang Enterprises, Inc. and Universidad Metropolitana.
The Green Bank Observatory is a facility of the National Science Foundation operated under cooperative agreement by Associated Universities, Inc.
A. Corsi acknowledges support from NSF Award \#1907975. DRL  acknowledges support from the NSF awards AAG-1616042, OIA-1458952 and PHY-1430284.  SBS acknowledges support from NSF grant AAG-1714897. SBS is a CIFAR Azrieli Global Scholar in the Gravity and the Extreme Universe program.

\section*{Data Availability}
The data underlying this article will be shared on reasonable request to the corresponding author.

\end{document}